# Reverse Engineering of Additively Manufactured Parts: Integrating 3D Scanning and Simulation-Driven Distortion Compensation

Jannatul Bushra[a], Md Habibor Rahman[a,b], Mohammed Shafae[a], Hannah D. Budinoff[a, 1]

[a] Department of Systems and Industrial Engineering, The University of Arizona, Tucson, AZ 85721, USA

[b] Department of Mechanical Engineering, University of Massachusetts Dartmouth, Dartmouth, MA 02747, USA

## ABSTRACT

**Purpose**: Reverse engineering can be used to derive a 3D model of an existing physical part when such a model is not readily available. For parts that will be fabricated with subtractive and formative manufacturing processes, existing reverse engineering techniques can be readily applied, but parts produced with additive manufacturing can present new challenges due to the high level of process-induced distortions and unique part attributes. This paper introduces an integrated 3D scanning and process simulation data-driven framework to compensate for distortions of reverse-engineered additively manufactured components. **Approach**: This framework employs iterative finite element simulations to predict geometric distortions and iteratively estimate the key dimensional characteristics of the part while accounting for process-induced distortion. The effectiveness of this approach is then demonstrated by reverse engineering two Inconel-718 components manufactured using laser powder bed fusion additive manufacturing. **Originality**: This paper presents a remanufacturing framework combining reverse engineering and additive manufacturing, leveraging geometric feature-based part compensation through process simulation, better capturing the design intent needed for reverse engineering. Our approach can generate both compensated STL and parametric CAD models, eliminating laborious experimentation during reverse engineering. We evaluate the merits of STL-based and CAD-based approaches by quantifying the accumulated errors induced at the different steps of the proposed approach and analyzing the impact of varying part geometries. **Findings**: Using the proposed CAD-based method, the average absolute percent error between simulation-predicted distorted dimensions and actual measured dimensions of the manufactured parts was 0.087%, with better accuracy than the STL-based method.

## 1 INTRODUCTION & BACKGROUND

Reverse engineering (RE) of part geometry for on-demand production using additive manufacturing (AM) enables lean operation when original part drawings or models are unavailable. This capability facilitates minimal-waste repair, avoids costly part inventories, and supports customized parts on demand (Raja and Fernandes, 2007), providing supply chain resilience. RE also supports creating digital twins without CAD models, reducing costs, minimizing errors, and accelerating the manufacturing process (Leng *et al.*, 2021; Xia *et al.*, 2023). This study focuses on geometric RE, which derives a 3D computer-aided design (CAD) model by extracting geometric information from a physical product (Anwer and Mathieu, 2016). The process includes capturing point clouds using 3D scanners, cleaning and merging multiple scans, and generating CAD models via surface fitting algorithms (Raja and Fernandes, 2007). Research has focused on improving scanning, data processing, and CAD reconstruction via improved techniques (Bénière *et al.*, 2013; Benkő *et al.*, 2001; Durupt *et al.*, 2011; Liu and Wang, 2011; Theologou *et al.*, 2015; Wells *et al.*, 2013). However, current RE approaches from academia and industry lack automated processes to account for the manufacturing-induced effects. While best practices suggest considering quality issues that may occur during the manufacturing process and adjusting the CAD model accordingly (Stratasys, 2018), there is a lack of systematic and automated methods to do so. In response, we propose a framework combining 3D scanning and AM distortion compensation as part of the RE process.

[1] Corresponding Author.
*E-mail address:* hdb@arizona.edu (Hannah D. Budinoff).

### 1.1 Reverse engineering for additive manufacturing

RE and AM complement each other (Elizondo and Reinert, 2019), and their combined application can shorten the product development cycle compared to using either in isolation (Kumar *et al.*, 2023; Milewski, 2017). The RE model obtained can be rapidly manufactured using AM processes for design validation or prototyping (Macy, 2015). RE can help create digital twins of AM parts, enabling process optimization (Cai *et al.*, 2020) and real-time monitoring (Pantelidakis *et al.*, 2022). The goal of many RE/AM research works has been to demonstrate the efficacy of a particular application (e.g., prostheses, masks, and car components) (Blaya *et al.*, 2018; Budinoff *et al.*, 2021; Juechter *et al.*, 2018; Ma *et al.*, 2018). However, the quality of remanufactured parts still needs to be systematically analyzed, as numerous factors can affect the quality of the produced model and remanufactured parts.

Figure 1 outlines our remanufacturing framework via a compensation-integrated-RE approach, from the manufactured to the remanufactured part, highlighting key errors at different stages. Manufacturing error $e_m$ occurs during part fabrication, resulting in dimensional deviations in the part. The AM manufacturing process introduces errors due to high thermal gradients and uneven shrinkage and cooling, which can vary depending on part geometry, process parameters, and build orientation (Bartlett *et al.*, 2018; Bushra and Budinoff, 2021). Scanning error $e_s$ may arise when capturing the manufactured part's geometry, with inaccuracies introduced by scanner resolution limits, and environmental conditions of the manufactured part (Boehler *et al.*, 2003). Specifically for AM, powder residues and reflective surfaces common in metal AM parts can cause issues (Wang and Feng, 2014). RE model construction error $e_{re}$ is introduced by geometric feature approximation and data filtering, smoothing, and interpolation when processing the original scan data (Rolls, 2001), which is especially problematic for the complex and intricate shapes that are common for AM (Javaid *et al.*, 2021). Model compensation error $e_{re.c}$ in generating a compensated model from the initial RE model occurs due to simplifications in material properties, boundary conditions, and mesh resolution (Sifakis and Barbic, 2012). Remanufacturing error $e_{rm}$ stems from the fabrication of the compensated model, where additional process-induced deviations occur. Finally, a scanning error can appear again during the final scan of the remanufactured part, denoted by $e_{rs}$, with limitations like those of the initial scan. Each error source throughout the RE process chain is interlinked, with deviations from earlier stages accumulating through the workflow and influencing subsequent steps. Therefore, an error-aware compensations approach is required to improve the dimensional accuracy in the RE process.

The relative inaccuracies of each step of the RE for the AM process were studied by (Bauer *et al.*, 2019) and (Elizondo and Reinert, 2019), but these works did not compensate for any manufacturing-induced distortion. Because of the complex distortion patterns and potential for significant distortion levels, especially for intricate parts (Shabani and Dukovski, 2022), CAD models planned for production using AM should be compensated to ensure the accuracy of the manufactured part. Effective geometric compensation relies on understanding, identifying, and accounting for the geometric inaccuracies in different stages of the RE process chain. Toward this goal, we identified and analyzed error sources at different stages of the RE workflow, from scanning inaccuracies in reverse-engineered models to process-induced geometric distortions. By accounting for the identified errors, more accurate compensation methods can be developed to enhance the fidelity and dimensional accuracy of remanufactured AM parts, thereby preserving the design intent in the part redesign.

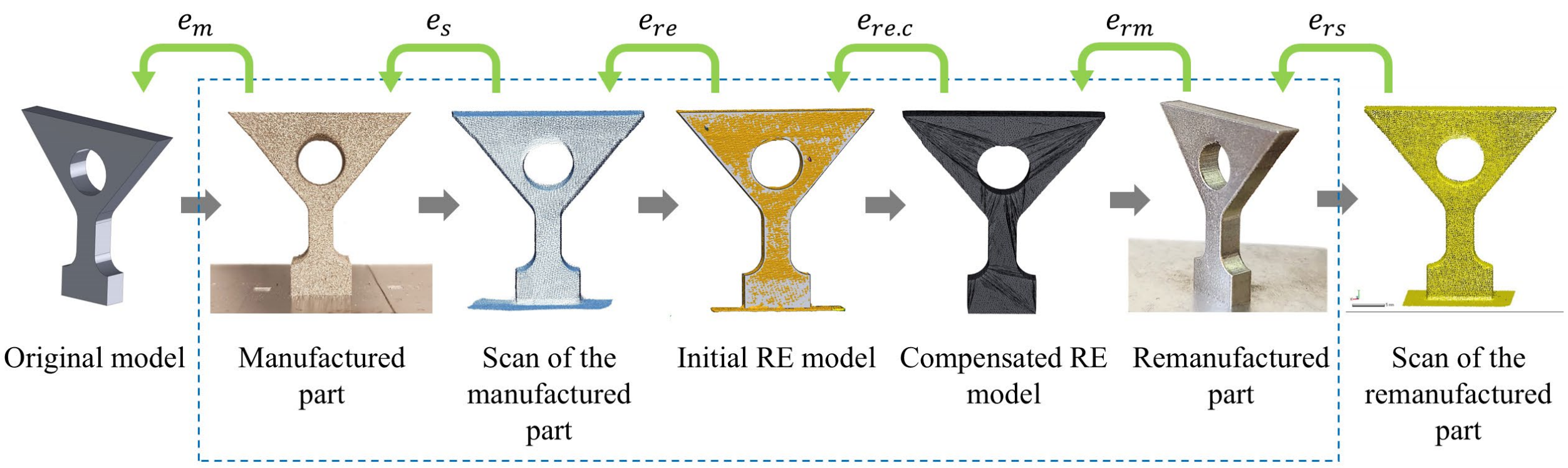

**Figure 1** Overview of our remanufacturing via the compensation-integrated-RE approach (shown within the dashed line boundary). Comparisons of the geometry from each step of the RE/remanufacturing process to the original model show how process steps introduce errors (e.g., $e_m, e_s, e_{re}$)

### 1.2 Geometric compensation in AM

Process-induced distortion for AM parts is caused by large thermal gradients resulting from rapid heating and solidification in AM processes (Denlinger and Michaleris, 2016; Hasan *et al.*, 2024; Ning *et al.*, 2019). This distortion causes warpage, dimensional inaccuracy, reduced tolerance, and compromised mechanical performance, ultimately impacting the quality and cost of the parts (Budinoff and Shafae, 2022; Denlinger, 2018; “Thermal distortion in AM”, 2024; Xie *et al.*, 2019). AM processes typically introduce non-uniform geometric distortions that vary in direction and across different geometrical features (Bartlett *et al.*, 2018; Bushra *et al.*, 2023, 2025). While many studies mitigate distortion by adjusting process parameters (e.g., scan speed, energy density) to reduce distortion (Denlinger, 2018; Hasan *et al.*, 2023; Xie *et al.*, 2019), our focus is on compensating the RE part geometry for AM process-induced distortions, prior to printing. To account for those distortions, the input part design should be compensated to result in the desired shape by identifying and inverting distortions at surface points or meshes relative to the desired geometric shape. Iteratively printing and adjusting CAD models to minimize distortion is expensive and time-consuming ("Distortion Prediction & Compensation - America Makes", 2024); simulation-based distortion prediction offers an alternative.

Researchers have performed part compensation studies for various AM processes, including directed energy deposition (Biegler *et al.*, 2020), binder jet (Paudel *et al.*, 2023), powder bed fusion (McConaha and Anand, 2020; Zhang *et al.*, 2020), polymer material extrusion (Wang *et al.*, 2021), and wire arc AM (Nguyen *et al.*, 2021) processes. However, these approaches often require time-consuming repeated manufacturing to identify appropriate compensation. More efficient simulation-based compensation is also a feature of several AM process simulation software tools (e.g., ANSYS Additive Print, Simufact Additive, Netfabb Local Simulation) to predict and compensate for the distortions in the manufactured parts by metal AM processes. AM process-induced distortions are highly variable due to machine-to-machine variation, material variety, complex part geometry, and variations in the AM process itself. The accuracy of the compensated designs is affected by the accuracy of those simulation tools (Peter *et al.*, 2020).

Current compensation approaches and tools also face challenges unique to RE applications. Most of these compensation approaches and tools return files in mesh (e.g., STL) format (Afazov et al., 2017; Biegler et al., 2020; Wang et al., 2021), which have a complex and rough surface, and are not easily editable in 3D CAD modeling software (Xiao and Roh, 2021). CAD models are needed for rapid and straightforward design changes. Converting the compensated mesh-format designs to parametric designs may cause additional errors in the design geometry (Xiao and Roh, 2021). Scanning-induced imperfections (e.g., holes, noise) further alter input geometry and hinder compensation convergence. To tackle this challenge, our proposed approach leverages geometric feature-based part compensation to improve the dimensional accuracy in remanufactured parts and preserve the design intent for redesign. This work lays the foundation for a unified RE–AM framework by integrating geometry reconstruction with simulation-driven compensation toward minimizing accumulated distortion across the remanufacturing process.

### 1.3 Research approach

The key challenges associated with existing RE approaches are:

1. Current RE approaches do not take manufacturing-induced distortion into account for remanufacturing purposes, which is problematic for AM parts.
2. AM compensation software tools typically return the compensated file in mesh format, which is not easily editable in 3D CAD modeling software, and converting mesh-format designs to parametric designs may introduce additional inaccuracies.
3. The accuracy of simulation tools to compensate RE models for the AM process has not yet been evaluated.

This paper addresses these challenges by proposing and evaluating an integrated approach to use simulation and scan data to compensate for AM process-induced distortions in the RE process. Our approach prioritizes compensating for key dimensional characteristics of a part as defined in a CAD file, making it easier for designers to comprehend, iterate, and implement. The differences between our geometry compensation approach in the proposed framework and the typical implementations in commercial software and existing literature are highlighted in Table 1. Measurement and design-related challenges in this process are discussed through case studies using powder bed fusion – laser-based (PBF-LB). The specific contributions of this work are:

1. An integrated RE-AM framework combining 3D scanning with iterative AM simulation-based distortion compensation (Sections 2.1-2.3).
2. Key dimensional characteristics-based compensation approach that addresses and preserves design intent both for RE and future redesign purposes (Section 2.2).
3. Demonstration of the use of the framework via conducting a case study (Sections 2.4-2.6).

**Table 1** Comparison of existing part compensation solutions

| Approach [category] | RE–AM workflow stages addressed | Output file | Output file is easily editable for RE | Provides clear guidance for automating iteration | User can identify important features to focus on |
|---|---|---|---|---|---|
| (“Ansys Additive”, 2021) [Commercial] | Distortion prediction & compensation | STL | No | No | No |
| (“Simufact Additive”, 2023) [Commercial] | Distortion prediction & compensation | STL and editable CAD model | Yes | Yes | No |
| (“Netfabb Simulation - Autodesk”, 2024) [Commercial] | Distortion prediction & compensation | STL | No | No | No |
| (“Simcenter 3D \| Siemens”, 2024) [Commercial] | Distortion prediction & compensation | Editable CAD model and STL | Yes | Yes | No |
| (Biegler *et al.*, 2020) [Academic] | Distortion compensation | Tool path | No | No | No |

| (McConaha and Anand, 2020) [Academic] | Distortion compensation | STL | No | No | No |
|---|---|---|---|---|---|
| (Zhang *et al.*, 2020) [Academic] | Distortion prediction (followed by NURBS surface fitting) | Editable CAD model | Yes | No | No |
| (Nguyen *et al.*, 2021) [Academic] | Distortion compensation | Tool path | No | No | No |
| Our CAD-based framework | Unified RE–AM pipeline: scanning, CAD reconstruction, simulation-driven compensation | Editable CAD model | Yes | Yes | Yes |

Our integrated framework bridges the traditionally separate stages of RE and AM by linking feature-based solid model reconstruction with simulation-driven distortion compensation. The identification and utilization of critical functional features guide both accurate parametric model reconstruction in RE and feature-based compensation in AM. Since scan data cannot be directly modified, identifying a parametric model via RE is essential to enable iterative adjustment of part geometry during AM compensation. The iterative adjustment of the RE part geometry through AM process simulations explicitly acknowledges the critical dependency between the fidelity of the initial RE model and the effectiveness of subsequent distortion compensation steps, thus forming a cohesive, distortion-aware RE-AM workflow. Current compensation approaches assume access to nominal CAD models, which is not possible for the RE use case explored here (which, by definition, has no nominal CAD model). Accordingly, a direct quantitative comparison with prior compensation studies is not feasible. We used one related RE and metal AM study (Bauer *et al.*, 2019) for benchmarking, which does not incorporate geometry compensation.

## 2 METHODOLOGY

We propose an RE and remanufacturing process (Figure 1) focusing on key dimensional characteristics-based part compensation to recreate components with minimal distortion when no CAD model exists. This process includes scanning the part, creating a CAD model, adjusting the CAD model to account for the specific AM process employed, and utilizing this compensated design for remanufacturing, as described in the following subsections.

### 2.1 3D scanning, key dimensional characteristics selection, and CAD model generation

Contact and non-contact metrology equipment can capture the part's shape by collecting point cloud or image data to generate the initial RE model, accurately digitizing the geometry of the as-manufactured part, especially when the original CAD model is unavailable. Manually generating a solid BREP directly from a physical part is challenging without complete surface information like that obtained by scanning. Data acquisition using 3D scanners involves multiple scanning passes covering different sides of the object to capture all geometric features (Elizondo and Reinert, 2019). Scanning metal AM parts brings unique challenges (e.g., high surface reflectivity, irregular surface from partially fused powder) that require adjusted scanner settings and surface preparation (e.g., matte spray) to improve accuracy. When 3D scanning cannot obtain reliable data, alternatives include CT scanning despite cost and accessibility barriers (Bauer *et al.*, 2019), multi-sensor fusion (Colosimo *et al.*, 2015), and advanced CAD model reconstruction (Zhang *et al.*, 2024).

Post-processing point cloud data often includes alignment, fusion, and filtering. Multiple scanned point clouds are aligned to a common coordinate system and then fused into a single dataset. Filtering techniques are applied to eliminate scanning noise. Next, the refined point cloud data is typically converted into a mesh structure. For AM parts, the rough surface of the printed part may require mesh cleaning and smoothing to fill small holes and isolated patches (Helle and Lemu, 2021).

When a CAD model is required, a model can be created by segmenting the mesh into distinct regions based on geometric features and curvature (e.g., planes, cylinders) to facilitate improved geometry replication (Benko *et al.*, 2001). Using mesh cross-sectional data, 2D sketches can be generated by projecting these sections onto defined planes. These sketches outline the object's geometry and help create individual geometric features through modeling operations like extrusions and cuts. During the sketching process, various geometric constraints (e.g., parallelism, equality) can be applied to lines and curves to ensure dimensional accuracy and maintain relationships between features. For AM parts, the fitting of regular geometric features to mesh data is especially important as prior work shows the inherent stairstep effect and STL discretization error result in an irregular part surface that was not present in the underlying CAD model (Klar *et al.*, 2019; Yanamandra *et al.*, 2020). Even with metal AM processes, some artifacts of the layered process and STL tessellation can be visible (Helle and Lemu, 2021). Careful scanning and thorough reconstruction – with a focus on removing artifacts from the AM process – are essential steps in our framework. Selection and measurement of key dimensional characteristics (KDCs) are directly extracted from the digitized geometry, ensuring a strong and direct linkage between scanning outcomes and subsequent CAD-based iterative compensation. Here, we define KDCs as product dimensional features (i.e., a part's overall size or the position or size of features) *that significantly impact the final cost, performance, or safety of a product when the key characteristics vary from nominal* (Thornton, 1999). Selecting KDCs can be challenging due to the lack of quantitative methods for their prioritization, but systematic approaches have been proposed (Idriss *et al.*, 2018; Thornton, 1999). These approaches selected KDCs whose statistical variability most strongly influences the system's non conformity rate (evaluated through global sensitivity analysis) (Idriss *et al.*, 2018), whose variation most significantly increases product quality loss (evaluated through variation sensitivity, process capability changes, and the cost–benefit of control) (Thornton, 1999).

## 2.2 Distortion prediction and compensation

Utilizing the initial RE model as input, our proposed framework employs iterative finite element simulations and 3D scanning to compensate for AM-induced distortion in KDCs. We introduce two new algorithms: one method compensates and returns a parametric RE-CAD model, and the other compensates and returns an RE-STL file. Our methods involve capturing KDCs of the part and utilizing AM simulation predicted distortions to assess the compensated part designs by iteratively adjusting the input 3D model (for the CAD-based compensation) or the compensation factor (for the STL-based compensation) based on those KDCs to account for the error between the actual dimensions and the dimensions from the simulations. The compensation factor is a multiplier applied to adjust a part's geometry based on predicted distortions, scaling the compensation of the geometry to pre-deform it in the opposite direction of manufacturing-induced expansion or contraction. Algorithms 1 and 2 show the steps involved in the iterative KDCs-based compensation process. The iterative compensation methods ensure that the resulting compensated CAD or STL model results in final parts closely aligned with the intended specifications. Then, the CAD-based and STL-based methods are compared.

Geometry compensation uses thermomechanical AM simulations to predict geometrical distortions due to high thermal gradients and uneven shrinkage and cooling. The software differs in the file format of the geometry used during the compensation process: some return compensated part files in CAD format (e.g., Simufact Additive), while some return them in STL format (e.g., Ansys Additive Print, Netfabb Local Simulation), and some provide them in both BREP and STL formats (e.g., Simcenter 3D for AM). The proposed framework involves running AM process simulations for the modified CAD/STL models found in different iterations of the compensation process. For example, we used ANSYS Additive Print simulations to predict distortions in the case studies. Studies have shown that ANSYS Additive Print can predict distortions in different test geometries with accuracy typically within ±50 µm of measured values (Jagatheeshkumar *et al.*, 2023; Mayer *et al.*, 2020) and excellent agreement between measured and modeled stresses (Wheeler *et al.*, 2022).

Error metrics for AM parts include achievable geometrical tolerances, dimensional error, and volumetric error (Rebaioli and Fassi, 2017). In both proposed compensation methods, several KDCs serve as reference points for capturing the design intent of the existing physical part during the compensation process and are used to minimize error on the most critical part features. While we demonstrate our approach here on single-parameter KDCs, many tolerances depend on multiple CAD parameters. Future implementations should integrate sensitivity analysis and multi-parameter optimization to address these complex cases. Another limitation is that our methods reduce accumulated deviations but do not yet quantify stage-specific error sources such as scanning, CAD generation, or fabrication. Future work should explicitly model error propagation to improve stage-aware compensation.

#### *2.2.1 Method 1: STL-based compensation*

In many compensation methods (Afazov *et al.*, 2017; "ANSYS Distortion Compensation", 2020; Shaikh *et al.*, 2021), the distortion compensation applied to the geometry is determined by a distortion compensation factor, which is scaled by the simulation-predicted distortion magnitude and applied to the original STL file. As the compensation factor giving the best-compensated geometry is highly dependent on the input geometry, the distortion compensation factor needs to be iterated over to improve the results ("ANSYS Distortion Compensation", 2020). ANSYS Additive suggests applying a compensation factor for the first iteration of the compensation process. The compensated file is used to rerun the simulation and observe whether the part is over- or under-compensated. Based on that, the scale factor is adjusted for the next iteration, and the compensation process is fine-tuned to achieve the desired outcome.

In our proposed algorithm, shown in the table below, the first iteration starts with the generated CAD model from 3D scanning as input geometry $G$ and initial compensation factor $CF_{1,j}, \forall j \in KDC$. For every iteration, we run two simulations to find the compensated model, $G_{CS}$, and then to predict the distortion of the compensated model using a pre-defined set of key dimensional characteristics, $KDC$. At each iteration, the dimensions of different KDCs ($j \in KDC$) in the compensated STL file ($G_{CS}$) are measured using point-to-point distance from the STL file. Multiple measurements are taken and averaged to reduce the impact of errors in the measurements resulting from the STL triangulation. From the second simulation result, based on the predicted direction of the distorted surface from the nominal part surface, we determine the extent of distortion experienced by each KDC using Equations (1) and (2), focusing on the linear dimensions of size measured from left to right of a feature or top to bottom of a feature. Here, $d_{ijk}$ denotes the $k^{th}$ measured distortion for a characteristic $j$ during iteration $i$. Then, we calculate the average of the $n$ sample distortion predictions on a specific surface. Here, $D_{l_{ij}}$ is the average distortion on the left or bottom surface, and $D_{r_{ij}}$ is the average distortion encountered on the right or top surface of the KDCs, assuming that the axis is directed from left/bottom to right/top.

$$D_{l_{ij}} = \frac{\sum d_{ijk}(l)}{n}, \text{ for } k = \{1,2,\dots,n\} \quad \text{-------- (1)}$$

$$D_{r_{ij}} = \frac{\sum d_{ijk}(r)}{n}, \text{ for } k = \{1,2,\dots,n\} \quad \text{-------- (2)}$$

$\forall j \in KDC$, we calculate the distorted dimension from the average distortion values found using equations (1) and (2) and the input STL dimension D. The equation to find the simulation-predicted distorted dimension of the KDC for the manufactured part is given by,

$$D_{S_{ij}} = D + D_{r_{ij}} - D_{l_{ij}} \quad \text{-------- (3)}$$

Using the compensated STL dimensions ($D_{CS_{ij}}$), simulation-predicted distorted dimensions ($D_{S_{ij}}$), actual measured dimensions ($D_{M_j}$) of the initial manufactured part corresponding to a key dimensional characteristic $j \in KDC$, and the compensation factor ($C_{F_{ij}}$), we calculate the adjusted compensation factors for each KDC using equation (4). To mitigate potential instability in Equation (4) when ($D_{CS_{ij}} - D_{M_j}$) ~ 0, the computed compensation factor $C_{F_{i+1,j}}$ was constrained within the range [–5, 5], and values outside this interval were reassigned to $CF_i$ , the compensation factor from the immediately preceding update. This range is provided by Ansys guidance ("Additive Print", 2021), based on typical values needed to compensate for a range of part geometries. The negative values within this range are used to reverse the distortion direction (i.e., when an expansion of a feature is predicted but the feature actually contracts).

$$C_{F_{i+1,j}} = \left(\frac{D_{CS_{ij}}}{D_{S_{ij}}} \times D_{M_j} - D_{M_j}\right) \times \frac{C_{F_{ij}}}{D_{CS_{ij}} - D_{M_j}} \qquad \text{-------- (4)}$$

If the absolute percent error (APE) (Eqn. 5) for all KDCs decreases, the average of those new compensation factors corresponding to all KDCs is used as the compensation factor in the next iteration. We continue iterating the same way until the APE for all KDCs reaches an APE threshold, $T_A$, as our stopping criteria.

$$APE_{ij} = \frac{\left|D_{S_{ij}} - D_{M_j}\right|}{D_{M_j}} \times 100\%, \;\; \forall i = \{1, 2, 3, \dots\}, j \in KDC \qquad \text{-------- (5)}$$

**Algorithm 1.** STL-based compensation approach

| | |
|---|---|
| **Input:** | Initial CAD model ($G$), APE threshold ($T_A$), max iteration ($N$), initial compensation factor ($C_{F_1}$) |
| **Output:** | Compensated STL file ($G_{CS}$) |
| **1** | Initialize iteration counter $i \leftarrow 1$, Input geometry $G \leftarrow$ Initial CAD model |
| **2** | **while** $i \leq N$ **do** |
| **3** | Perform FEM simulation on $G$ and generate the compensated STL file $G_{CS_i}$ using $C_{F_i}$ |
| **4** | Rerun the FEM simulation on $G_{CS_i}$ to predict process-induced distortion |
| **5** | Measure the input key dimensional characteristic from $G_{CS_i}$ |
| **6** | **for each** characteristic dimension $j \in KDC$ (the set of KDCs) **do** |
| **7** | Find the simulation-predicted distorted dimension $D_{S_{ij}}$ using equation (3) |
| **8** | Compensation factor for $j$, $C_{F_{i,j}} \leftarrow C_{F_i}$ |
| **9** | Calculate the new compensation factor ($C_{F_{i+1,j}}$) using equation (4) |
| **10** | Calculate $APE_{ij}$ using equation (5) |
| **11** | **if** $i \geq 2$ and $APE_{ij} > APE_{i-1,j}$ |
| **12** | Compensated STL, $G_{CS} \leftarrow G_{CS_i}$ |
| **13** | **end if** |
| **14** | **end for** |
| **15** | **if** $APE_{ij} \leq T_A, \forall j \in KDC$ |
| **16** | Compensated STL, $G_{CS} \leftarrow G_{CS_i}$ |
| **17** | **return** < Compensated STL file, $G_{CS}$ > |
| **18** | **else** $C_{F_{i+1}} = \sum_j C_{F_{i+1,j}} / \sum_j 1$ |
| **19** | $i \leftarrow i + 1$ |
| **20** | **end if** |
| **21** | **end while** |
| **22** | **return** < Compensated STL file, $G_{CS}$ > |

#### 2.2.2 *Method 2: CAD-based compensation*

In our second algorithm, we aim to find the compensated input CAD model for an AM process that will result in a final product with minimal deviation from the existing part. The compensated file is a parametric CAD model, which is helpful when further modification or update is needed. In the first iteration, we feed the generated CAD design from 3D scanning into the AM simulation. From the simulation results, for every key dimensional characteristic $j \in KDC$, we calculate the simulation-predicted distorted dimension following the same procedure described in the previous section using equations (1), (2), and (3). The only difference is that the input dimensions for each iteration are the dimensions of the KDC of the input CAD in that iteration.

Next, we compare the simulation-predicted distorted dimensions $D_{S_{ij}}$ with the actual measured dimensions $D_{M_j}$ of the KDC and calculate the error ($\epsilon$) between them using equation (6).

$$\epsilon_{ij} = D_{S_{ij}} - D_{M_j} \qquad \text{---------} \ (6)$$

We adjust the new input CAD dimensions $D_{S_{i+1,j}}$ for the next iteration or simulation using equation (7) based on the error $\epsilon$ and the scale factor $S_F$.

$$D_{S_{i+1,j}} = D_{S_{ij}} - \epsilon_{ij} \times S_F \qquad \text{---------} \ (7)$$

Here, $S_F$ is the scale factor used to scale the error and adjust the CAD dimension for better results. We repeat this modification process iteratively, updating all KDCs' dimensions until the APE for each KDC falls below a specified threshold.

**Algorithm 2.** CAD-based compensation approach

| | |
|---|---|
| **Input:** | Initial CAD model, APE threshold ($T_A$), max iteration ($N$), scale factor ($S_F$) |
| **Output:** | Compensated CAD design ($G_{CC}$) |
| **1** | Initialize iteration counter $i \leftarrow 1$, Input geometry $G_1 \leftarrow$ Initial CAD model |
| **2** | **while** $i \leq N$ **do** |
| **3** |     Perform FEM simulation to predict the process-induced distortion on $G_i$ |
| **4** |     **for each** characteristic dimension $j \in KDC$ (the set of KDCs): |
| **5** |         Find the simulation-predicted distorted dimension $D_{S_{ij}}$ using equation (3) |
| **6** |         Calculate the error $\epsilon_{ij}$ using equation (6) |
| **7** |         Find new input CAD dimension from equation (7) |
| **8** |         Calculate $APE_{ij}$ using equation (5) |
| **9** |         **if** $i \geq 2$ and $APE_{ij} > APE_{i-1,j}$ |
| **10** |             Compensated CAD design $G_{CC} \leftarrow G_i$ |
| **11** |             **return** < Compensated CAD design, $G_{CC}$ > |
| **12** |         **end if** |
| **13** |     **end for** |
| **14** |     **if** $APE_u \leq T_A, \forall u \in KDC$ |
| **15** |         Compensated CAD design $G_{CC} \leftarrow G_i$ |
| **16** |         **return** < Compensated CAD design, $G_{CC}$ > |
| **17** |     **else** Reconstruct CAD using updated dimensions $D_{S_{i+1,j}}$ |
| **18** |         $i \leftarrow i + 1$ |
| **19** |     **end if** |
| **20** | **end while** |
| **21** | **return** < Compensated CAD design, $G_{CC}$ > |

The proposed compensation approaches aim to reduce the total accumulated deviation between the original and remanufactured parts by compensating for the simulation-predicted distortions. However, the current framework does not quantify the contribution of individual errors introduced during specific RE stages such as scanning, CAD generation, or fabrication. This work lays the groundwork for future modeling efforts to analyze error propagation and implement stage-aware compensation strategies.

### 2.3 Comparison metrics

The final step in the RE framework is to remanufacture the compensated part design using the same process parameters used in the part compensation process. To assess the accuracy of the distortion compensation methods, the compensation errors of the compensated model dimensions compared to the original model dimension for each KDC, $j \in KDC$ are calculated as follows for STL and CAD models, respectively:

$$\epsilon_{CS_j} = \left| D_{CS_j} - D_j \right|, \forall j \in KDC$$
$$\epsilon_{CC_j} = \left| D_{CC_j} - D_j \right|, \forall j \in KDC$$

Additionally, to evaluate error propagation through each process step for the entire part, rather than just the selected KDCs, using an inspection software, we perform best-fit alignment comparison between the outputs (i.e., scans or models) generated at various stages within our framework and the original model, obtain the deviation distributions between the outputs and the original model, and calculate the mean and standard deviation from the distortion distributions, following the approach of (Bauer *et al.*, 2019). This comparison helps us understand the error between the input (e.g., original model) and the desired output (e.g., scan of a remanufactured part), analyzing errors among different steps.

## 2.4 Case study – components, materials, and equipment

Following our proposed methods, we conducted two case studies to evaluate the feasibility of remanufacturing a RE-component using AM. The case study parts are: (i) an overhang test specimen with a bounding box size of $30mm\ x\ 4mm\ x\ 30mm$, and (ii) a perforated thin plate of $48mm\ x\ 2mm\ x\ 53mm$ dimension (Figure 2).

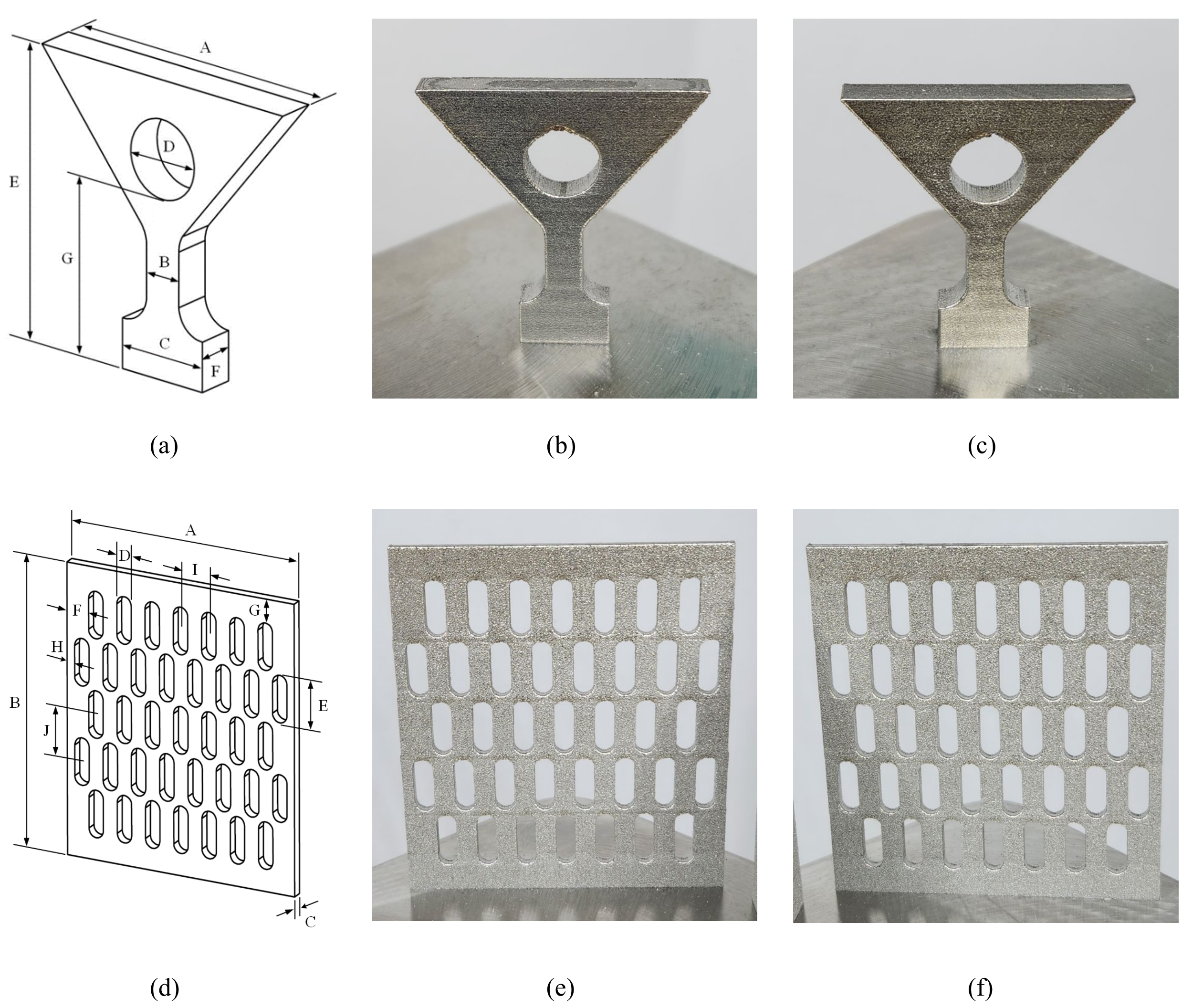

**Figure 2** KDC of the overhang test specimen (a) and the thin perforated plate (d), with the corresponding remanufactured specimens using the STL-based method (b, e) and CAD-based method (c, f)

These test components were selected to have a variety of KDC types and geometrical features. We also established knowledge of those components concerning their manufacturing process, parameters, materials, and original CAD designs. We manufactured the parts using a Concept Laser (Mlab Cusing R) machine and Inconel-718 material. The process parameter settings include $25^{\circ}\mathrm{C}$ baseplate temperature, $95\ \mathrm{W}$ laser power, $500\ \mathrm{mm/s}$ scan speed, $110\ \mu\mathrm{m}$ hatch spacing, $15\ \mu\mathrm{m}$ layer thickness, $0^{\circ}$ and $90^{\circ}$ starting layer angle (overhang test specimen and perforated thin plate, respectively), $90^{\circ}$ and $180^{\circ}$ layer rotation angle (overhang test specimen and perforated thin plate, respectively). The same parameters were used to remanufacture the compensated 3D models derived from the two distortion compensation methods.

## 2.5 Case study - 3D scanning and data processing

We used the FARO Quantum S ScanArm with a Laser Line Probe to collect 3D point cloud data. Then, Geomagic Control X Software was used to post-process this data, performing alignment, fusion, and filtering. Multiple point clouds obtained from several scanning passes covering different sides of the object were aligned. The aligned point clouds were combined into a single point cloud and post-processed to eliminate outliers introduced during the scanning process. Next, we utilized Geomagic Design X Software to generate parametric CAD models from scan data. We converted the point cloud data from the scan into a mesh structure and utilized the mesh to construct the CAD model. The process involved automatically segmenting the mesh into distinct regions, defining coordinate planes, generating 2D sketches, and employing these sketches to create individual geometric features.

Segmentation was performed using the "Auto Segment" function, which classified mesh data into regions such as planes and cylinders for both parts—utilizing curvature and features—for improved replication. The mesh structure was then oriented using the "Interactive Alignment" function, leveraging the newly classified regions. The "Mesh Sketch" tool projects a cross-section on a plane for sketching the outlines. Finally, the solid body was generated using the "Extrude" function. Extrudes and cuts were used to create the CAD model for both parts. When creating the sketches of the part's cross-section, we applied parallel, perpendicular, and equal constraints to several lines. For example, for the overhang test specimen, the two fillet radii on the left side of the part are assumed to match the two fillet radii on the right, and the top and bottom surfaces are considered parallel. For the perforated thin plate, the slot dimensions were assumed to be equal, and we applied the average dimensions of all the slots as the slot dimensions; the sides were assumed to be parallel to each other and perpendicular to the bottom flat surface.

## 2.6 Case study – process simulation, simulation calibration, and geometry compensation

We utilized the compensation feature of ANSYS Additive Print thermal strain (TS) simulation software and the on-plate distortion prediction feature. The TS simulation utilizes thermal and mechanics solvers to model the periodic heating and rapid cooling from the PBF-LB process, as well as the subsequent shrinkage that induces deformation. In the thermal solver, the bottom of the build is constrained with a fixed-temperature boundary condition equal to the user-specified baseplate temperature, a Gaussian laser heat source is applied, uniform forced convection is imposed on the top surface, radiation effects are neglected, and the lateral boundaries are treated as adiabatic with a surrounding powder buffer (“Additive Print”, 2021). In the mechanical solver, the bottom of the build is fully constrained (“Additive Print”, 2021). Assumptions such as these are common in component-scale LPBF simulations as they reduce computational time while still providing good agreement with experimental results (Afazov *et al.*, 2022; Khanbolouki *et al.*, 2024). We used the same process parameters set for the initial part printing process, Inconel-718 properties, and the selected build orientation in the simulation. We used the standard Ansys pre-defined Inconel-718, with elastoplastic material behavior based on the J2 (von Mises) plasticity model and temperature- and state-dependent material properties (e.g., density, conductivity, specific heat) (“Additive Print”, 2021). The proprietary AM-specific material properties have been developed to match experimental results and shown to have good predictive capabilities (“Additive Print”, 2021; Jagatheeshkumar *et al.*, 2023; Mayer *et al.*, 2020; Wheeler *et al.*, 2022). Similarly to prior work, the effect of creep is not modeled here (Khanbolouki *et al.*, 2024). To ensure the best possible fit to experimental results, we have also experimentally determined the calibration factors to account for the impact of material-machine interactions on the development of strains in the part.

We performed a voxel sensitivity study to balance accuracy and computational cost using a part with overhanging features of the same approximate size and complexity as the studied parts. We varied the voxel size and the voxel sample rate (i.e., rate at which each voxel is subdivided) from 0.2 to 0.8 and 2 to 10, respectively, and found that distortion predictions stabilized at a voxel size of 0.3 mm and a voxel sample rate of 6; these are the values adopted here. We calibrated the simulations for the machine and material used in printing these parts, following the process recommended by ANSYS ("Additive Calibration", 2021) to determine calibration factors denoted as Strain Scaling Factors and Anisotropic Strain Coefficients. The calibration factors are valid for a particular printed geometry and geometries with similar features. We used a standard ANSYS calibration part with similar size and features to our case study parts (e.g., thin walls, slanted, and overhanging faces) to ensure the calibration factors were appropriate ("Additive Calibration", 2021). The following calibration factors were obtained through iterative calibration: strain scaling factor: 4.685, anisotropic strain coefficients parallel to scan direction: 0.743, anisotropic strain coefficients perpendicular to scan direction: 1.257, anisotropic strain coefficients along build direction: 1. Following the voxel sensitivity results, for the simulation calibration and the geometry compensation, the voxel size was set to 0.3 mm the voxel sample rate was 6, and the mesh resolution factor was 4. We used elastic-plastic material behavior and the J2 (Von Mises) plasticity model to capture the deformation more accurately. Each simulation took 5 to 6.5 hours on an eight-core Intel® Core™ i7-10700 processor. For the distortion compensation algorithms, we utilized the APE threshold, $T_A$ of 0.5%. We set $S_F = 0.75$ and $CF_{1,j} = 0.75 \, \forall j \in KDC$, as ANSYS Additive suggests a compensation factor of 0.75 for the first iteration of the compensation process.

# 3 RESULTS

The case study aimed to quantify the errors between the original and compensated models (Table 2 and Table 3) and the manufactured parts and final remanufactured parts (Figure 2 and Figure 4). In the initial RE model generation step, the produced RE models were relatively accurate, with deviations between the mesh scan data and the initial RE model falling between $\pm$ 0.03 mm. Some areas of higher deviation were present, namely in overhanging surfaces and at 90° edges between flat surfaces for the overhang test specimen and internal surfaces of the slots and around 90° edges between flat surfaces for the perforated thin plate. We used the two methods discussed in Section 0 to compensate for the initial RE model for the AM process. We performed two iterations for the STL-based compensation method for both parts, iteratively changing the compensation factor. The final distortion compensation factor was 1.019 for the overhang test specimen and 0.891 for the perforated thin plate. The stopping criteria for the CAD-based compensation method were reached in three iterations for both parts. For the CAD-based compensation approach, the APE between the simulated predicted dimension and the measured average dimension of the initial manufactured part (i.e., initial RE part dimension) decreases for all seven KDCs of the overhang test part (Figure 3). With the STL-based compensation approach, the APE decreases for some KDCs while increasing for others. For all KDCs in both parts, the average APE was 1.82% for the STL-based method and 0.087% for the CAD-based method.

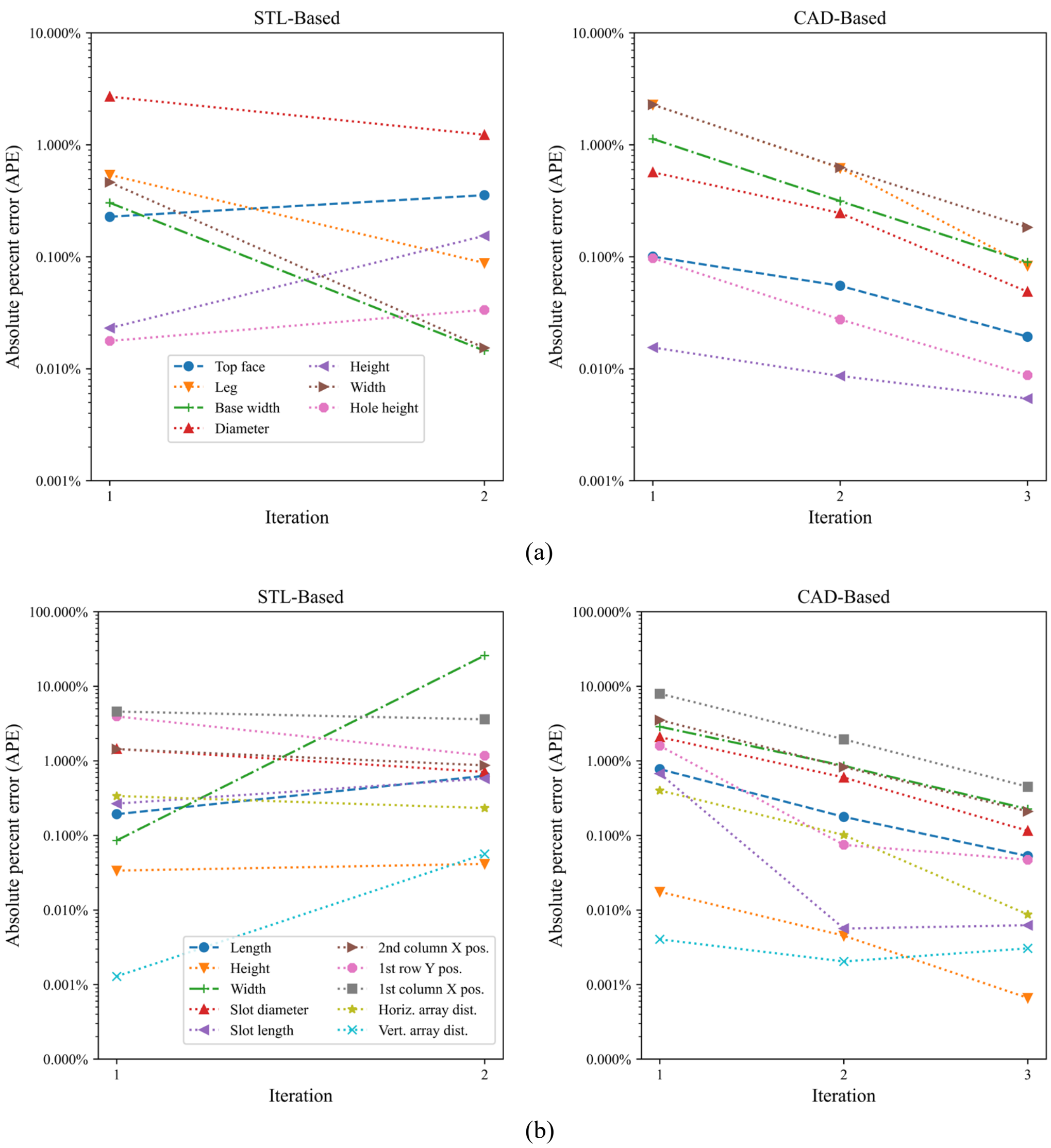

**Figure 3** APE for the KDCs of (a) the overhang test specimen and (b) the perforated thin plate over iterations

## 3.1 Compensation error for the overhang test specimen

The error was quantified for the KDCs to explore the agreement between the models (Table 2). For the STL-based method, we report the compensated model dimensions as the mean ± standard deviation, calculated from six repeated measurements ($n = 6$) per feature. These measurements were taken along the length of each KDC using the SolidWorks measurement tool, with points distributed to provide representative coverage of the mesh facets corresponding to that feature. This approach captures the variability introduced by mesh discretization and facet selection. In contrast, the CAD-based dimensions exhibit no variation, as they are defined deterministically in the reconstructed parametric model. The maximum compensation error for the CAD-based method is relatively small, measuring around 0.195 mm for the part height (E). However, the maximum error of the STL-based method is higher, 0.255 mm for the hole diameter (D). The magnitude of compensation error fluctuates based on the part feature being considered. The compensation error is less for the CAD-based compensation method for all KDCs, except for the top face (A).

**Table 2** Key dimensional characteristics and compensation error for overhang test specimen

| Key dimensional characteristics | Original model nominal dimension [mm] | Compensated STL model dimension (mean ± SD) [mm] | Mean compensation error for STL-based, $\epsilon_{CS}$ [mm] | Compensated CAD model dimension [mm] | Mean compensation error for CAD-based, $\epsilon_{CC}$ [mm] |
|---|---|---|---|---|---|
| Top face (A) | 30.00 | 29.948 ± 0.008 | 0.052 | 29.855 | 0.145 |
| Leg (B) | 4.00 | 4.071 ± 0.006 | 0.071 | 4.067 | 0.067 |
| Base width (C) | 10.00 | 10.063 ± 0.009 | 0.063 | 10.058 | 0.058 |
| Diameter (D) | 8.00 | 7.745 ± 0.136 | 0.255 | 7.883 | 0.117 |
| Height (E) | 30.00 | 29.762 ± 0.056 | 0.238 | 29.805 | 0.195 |
| Width (F) | 4.00 | 4.043 ± 0.007 | 0.043 | 4.041 | 0.041 |
| Hole height (G) | 18.00 | 17.963 ± 0.002 | 0.037 | 17.967 | 0.033 |
| Average compensation error | | | 0.108 | | **0.094** |

Figure 4 (a-c) compares the initial scan of the manufactured part and the scan of the remanufactured part, obtained from STL- and CAD-based methods. Both scans were aligned with the original model to facilitate this comparison. The scan of the remanufactured part obtained from CAD-based compensation exhibits a closer match to the scan of the manufactured part compared to the scan derived from STL-based compensation. However, both methods have weaknesses. The STL-based method struggles with overhang surfaces, curved features (i.e., fillets and hole dimensions), and part thicknesses. While the CAD-based method performs better overall, it still faces challenges with fillet and hole dimensions, though less than the STL-based method.

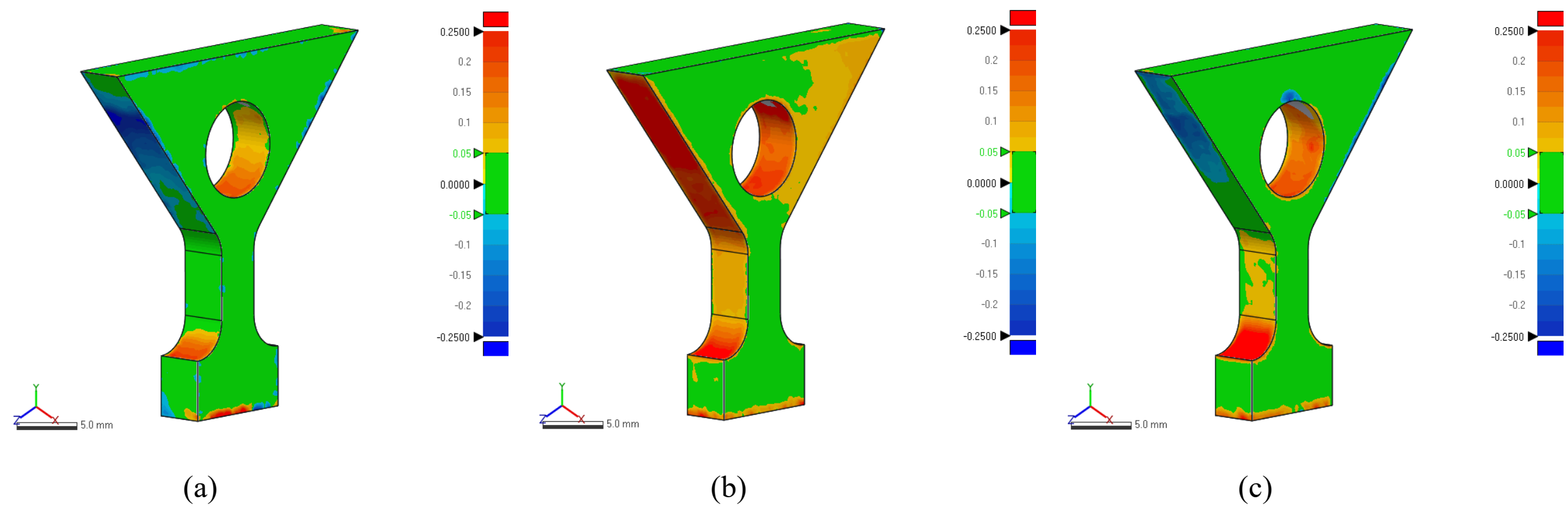

(a) (b) (c)

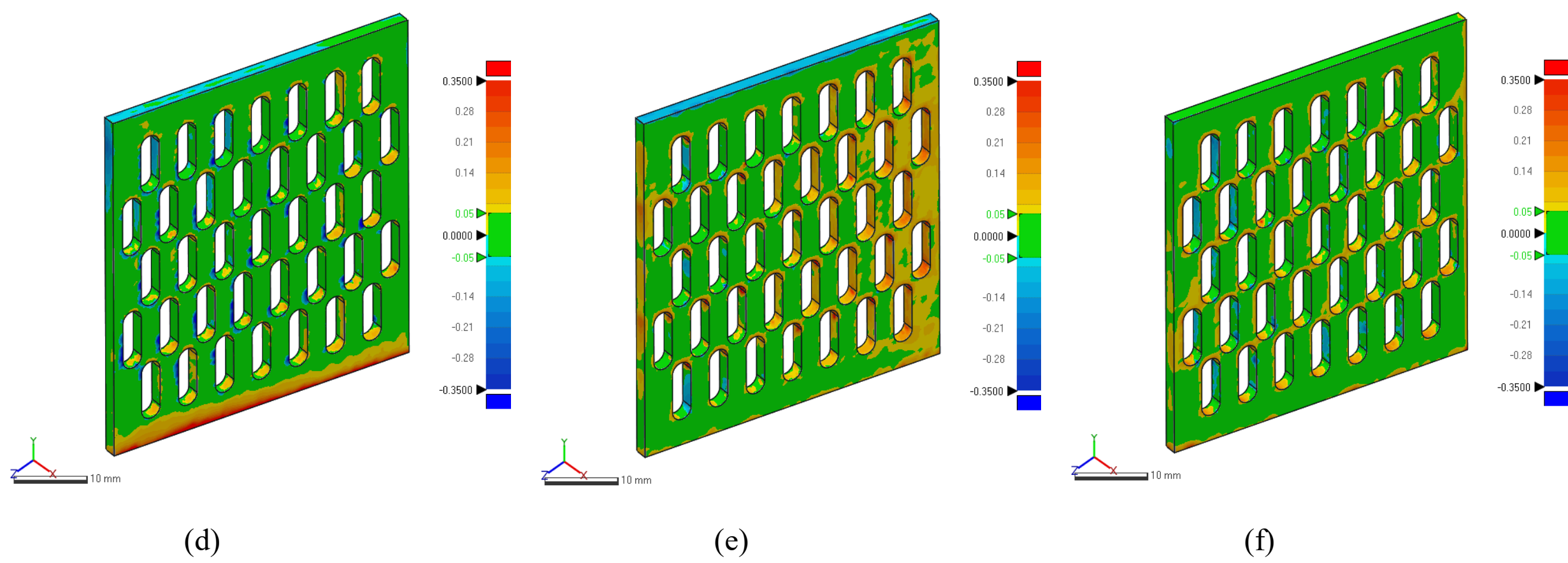

(d) (e) (f)

**Figure 4** Best fit alignment of the original model and scans of (a, d) manufactured part, (b, e) remanufactured part produced using STL method, (c, f) remanufactured part produced using CAD method for the overhang test specimen and the perforated plate. The color scale units are mm.

## 3.2 Compensation error for the perforated thin plate

Table 3 compares the digital models (i.e., the original and compensated RE models). The error was quantified for each KDC to explore the agreement between the digital models. The maximum compensation error for the CAD-based compensation method is around 0.463 mm for slot length (E), but for the STL-based compensation method, the maximum error is higher (0.561 mm for part width (C)). The CAD-based compensation method exhibits lower error for specific KDCs, while the STL-based compensation method performs better for others. However, on average, the CAD-based compensation method demonstrates lower compensation error.

Figure 4 (d-f) compares the initial scan of the manufactured part and the scan of the remanufactured part, obtained from both STL- and CAD-based methods. The scan of the remanufactured part from CAD-based compensation closely matches the manufactured part scan, outperforming the STL-based compensation method. However, both compensation methods exhibit shortcomings in certain aspects. The STL-based method performs poorly with the top and side surfaces, slot dimensions, and part thickness, whereas the CAD-based method performs better but still has issues with the top, bottom, and side surfaces.

**Table 3** Key dimensional characteristics and compensation error for thin perforated plate

| Key dimensional characteristics | Original model nominal dimension [mm] | Compensated STL model dimension (mean ± SD) [mm] | Mean compensation error for STL-based, $\epsilon_{CS}$ [mm] | Compensated CAD model dimension [mm] | Mean compensation error for CAD-based, $\epsilon_{CC}$ [mm] |
|---|---|---|---|---|---|
| Length (A) | 48.00 | 48.003 ± 0.011 | 0.003 | 48.303 | 0.303 |
| Height (B) | 53.00 | 52.928 ± 0.012 | 0.072 | 52.949 | 0.051 |
| Width (C) | 2.00 | 2.561 ± 0.005 | 0.561 | 2.046 | 0.046 |
| Slot diameter (D) | 3.00 | 2.995 ± 0.016 | 0.005 | 2.980 | 0.020 |
| Slot length (E) | 9.00 | 8.479 ± 0.009 | 0.521 | 8.537 | 0.463 |
| Second column X position (F) | 4.50 | 4.658 ± 0.012 | 0.158 | 4.698 | 0.198 |
| First row Y position (G) | 4.00 | 4.478 ± 0.009 | 0.478 | 4.424 | 0.424 |
| First column X position (H) | 1.50 | 1.629 ± 0.022 | 0.129 | 1.697 | 0.197 |
| Horizontal distance of the array (I) | 6.00 | 5.980 ± 0.026 | 0.020 | 5.993 | 0.007 |
| Vertical distance of the array (J) | 9.00 | 8.993 ± 0.008 | 0.007 | 8.990 | 0.010 |
| Average compensation error | | | 0.195 | | **0.172** |

### 3.3 Error propagation through the RE framework

To investigate the relative impact of each step in the RE and remanufacturing framework, we compared the overall geometric deviations of the parts or models to the same reference geometry—the original model. We aligned point clouds from different steps in the framework with the original model in Geomagic Control X software, enabling the extraction of average and standard deviation from distortion distributions and assessing error propagation through each step (Figure 5). We directly utilized scan data for the parts' point clouds. For the RE model, we generated point clouds by sampling points on the surfaces of the STL files for comparisons with the original model. The overall deviation between the point clouds and the original model for the whole part was visualized in Control X using the 3D Compare tool. The deviation represents the distance between the scan points and their projected points on the CAD or STL surfaces. From the statistical information created by the 3D Compare tool, we extracted the mean and standard deviation of the deviation distributions between the output of different steps and the original model. The standard deviation is higher in the steps involving part scanning. The standard deviation is also higher for the STL-based compensated RE model than the CAD-based compensated RE model for the overhang test specimen.

We observe notable differences in process-induced distortion and scanning-related errors for the case study parts. The overhang test specimen exhibits a lower overall deviation between the scan of the manufactured part and the original model, with mean deviation values close to zero and lower standard deviations, indicating reduced distortion and scanning error. In contrast, the perforated thin plate appears to have undergone overall expansion during printing, whereas the overhang specimen contracted—assuming minimal contribution from scanning error. Thin and intricate features in the perforated plate likely contributed to more significant initial process-related and scanning-induced distortions. In the initial RE models, the deviation between the RE and original models shows a downward shift relative to the deviation observed between the scan of the manufactured part and the original model. This shift increases the mean deviation in the overhang specimen, which had originally contracted. In contrast, the RE model for the perforated thin plate closely matches the original model, resulting in a mean deviation near zero—consistent with the fact that the manufactured part experienced expansion rather than contraction.

Comparing the two compensation methods, the CAD-based approach results in a lower mean and standard deviation of deviations between the compensated RE model and the original model for the overhang test specimen. However, for the perforated thin plate, the CAD-based method yields a slightly higher mean deviation between the compensated RE model and the original model than the STL-based method. For the CAD-based compensation, the mean deviation of the scan of the remanufactured part is closer to that of the scan of the original manufactured part, indicating better alignment with the original process outcome. The standard deviation of deviations between the scans of the remanufactured and original manufactured parts is approximately the same for both compensation methods.

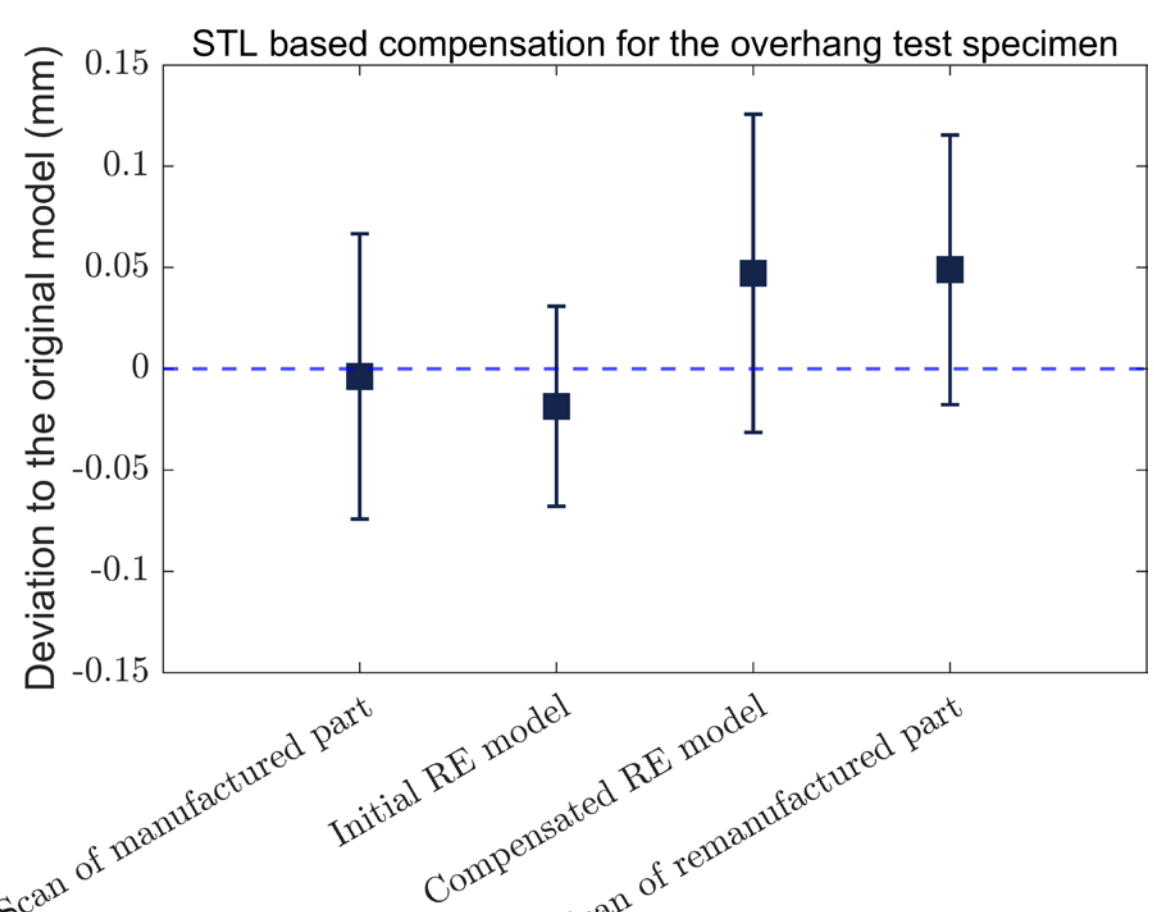

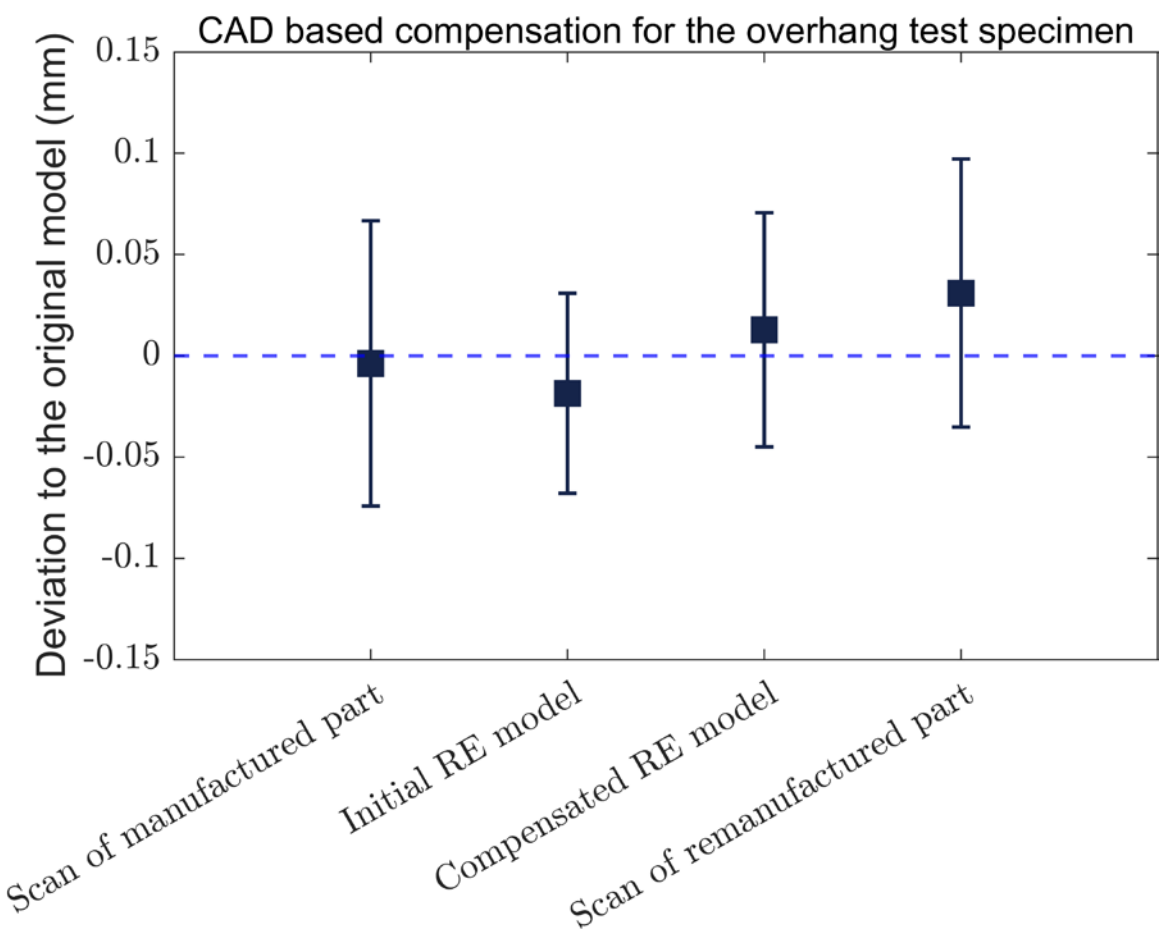

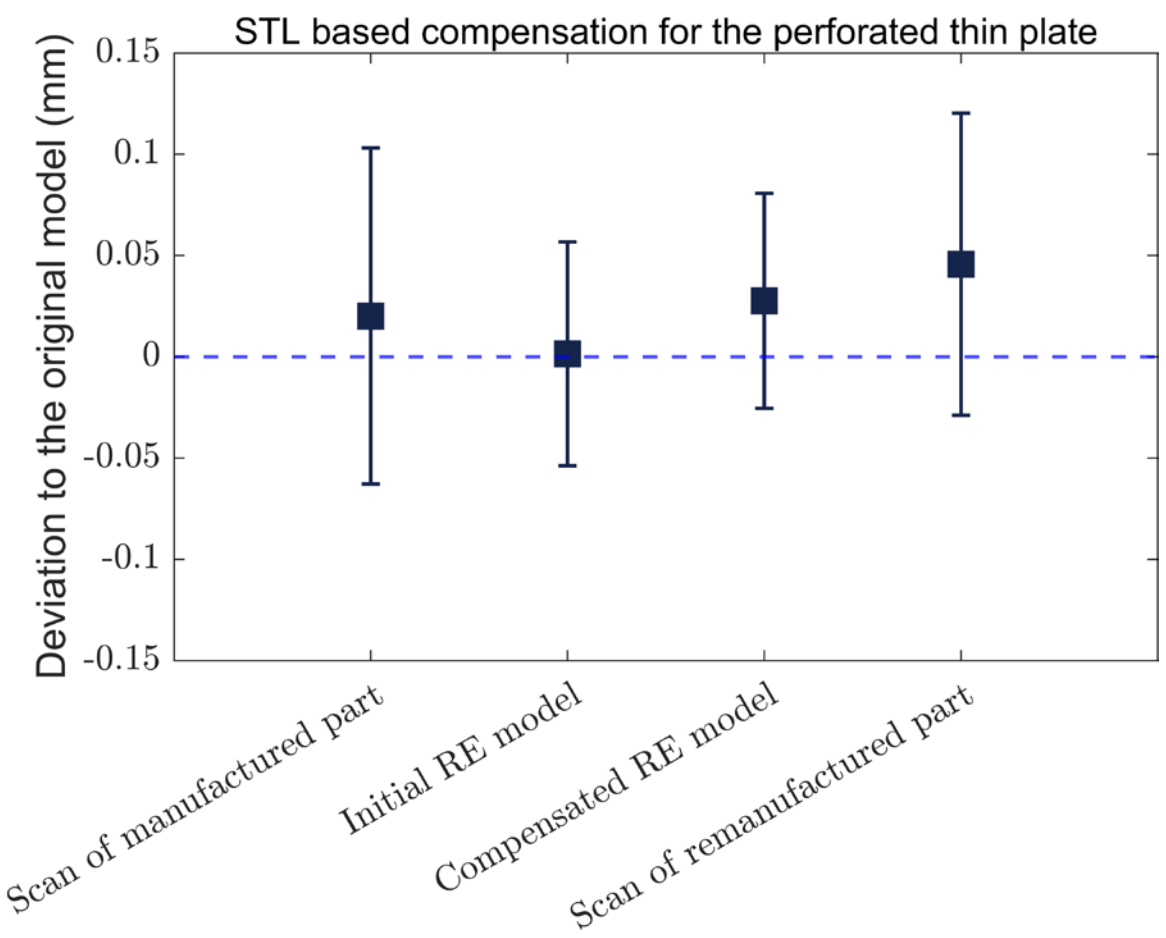

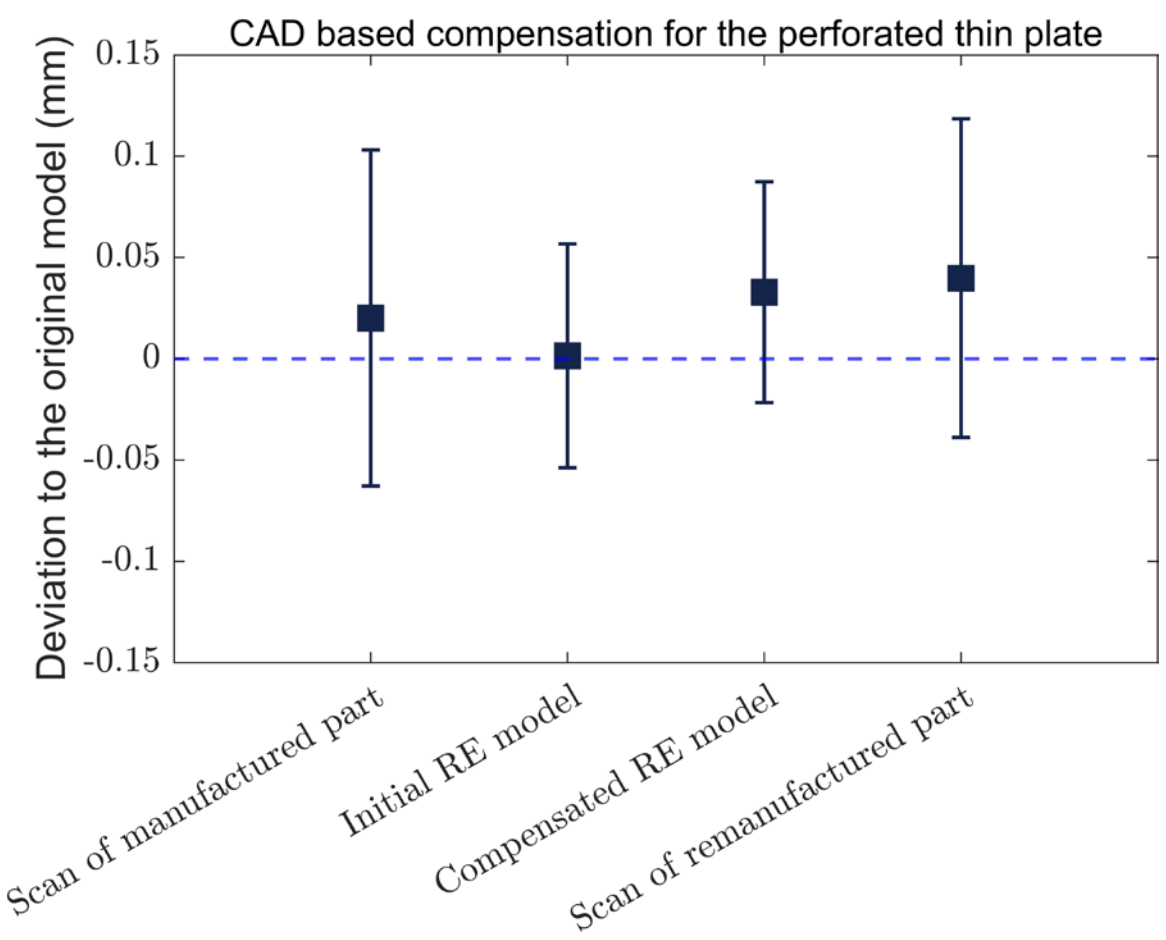

**Figure 5** Distributions (mean and standard deviation) of deviations introduced by each step of the RE framework from the original model.

This comparison shows the varied sources of error throughout each step of the RE process. The deviation distribution for scanning reflects manufacturing process-induced deviations and errors caused by the scanning process (e.g., poor lighting conditions and reflectivity). During RE, aligning and registering multiple scan data and surface fitting to meshes can introduce errors due to poor approximation of complex surfaces, but can also compensate for some process-induced distortion (e.g., use a perfectly flat face instead of the warped surface produced during manufacturing). During simulation, errors can be introduced by the uncertainty associated with the process parameters and model inaccuracies. However, the remanufactured parts are fairly accurate, with most points falling within 0.1 mm of the original model.

# 4 DISCUSSION

## 4.1 Sources of error in RE and remanufacturing

Our results help evaluate the effectiveness of compensation in improving RE for AM. The comparison between the scan of the original manufactured part and the original model (Figure 4a, Figure 4d, and Figure 5) reveals deviations, likely due to process-induced distortions and/or scanning errors. The initial RE model also deviates from the scan of the original manufactured part, highlighting the presence of distortions and errors during the CAD generation process (Figure 5). A similar finding is evident from the results of Elizondo and Reinert (2019) on comparing different RE modeling methods, indicating that parametric feature modeling improves surface flatness and smoothed welded areas with finer detail, though it results in a slight loss of accuracy when parametrizing the surfaces, while mesh modeling offers quicker but less refined surfaces (Elizondo and Reinert, 2019). Moreover, measurement errors are likely present, which affect the following steps in the remanufacturing process and propagate the errors further.

Scanning-related errors are significant for smaller parts because they impact the smaller dimensions more. The scans often have issues, such as noise and missing parts, especially in holes and slots on inner surfaces. Additionally, the rough surfaces and edges created by the AM process contribute to noisy scans, resulting in increased variability. Future work can improve accuracy for AM parts, as current noisy data can make RE and compensation difficult. Here, we matched the process parameters and orientation used for the original manufactured part throughout the RE and remanufacturing processes. Therefore, future work should explore the accuracy and uncertainty in compensation when the remanufacturing process parameters differ. Future work can also improve alignment with the initial RE model using the simulation-predicted distorted geometry, similar to material extrusion approaches that use stair step error estimates to improve scan alignment (Yang *et al.*, 2022).

### 4.2 Comparison with existing RE and remanufacturing approaches

The results presented in this work can be compared to existing RE/AM applications. The approach of Bauer et al. (2019) involved reverse engineering a turbine blade by scanning an additively manufactured test part with optical and CT scans and then reproducing it through additive manufacturing (Bauer *et al.*, 2019). Their remanufacturing framework differs from ours, as we generate a CAD model from the part scan data and apply geometry compensation before printing, whereas they directly print the reverse-engineered mesh without optimizations such as surface smoothing or geometry compensation. From their results, the deviation distributions of the regions of interest in the additively manufactured replica compared to the nominal CAD geometry have mean values ranging from -35 µm to 61 µm and standard deviations ranging from 60 µm to 136 µm. Whereas, for our case, the deviation distributions of the four remanufactured parts compared to their corresponding nominal CAD geometries have mean values ranging from 30.0 µm to 48.9 µm and standard deviations ranging from 66.2 µm to 78.6 µm. Since our methods differ, we cannot definitively attribute the accuracy difference to any specific factor. However, additional steps for the RE CAD model generation and geometry compensation in the remanufacturing framework could be a potential reason for the difference in results.

Future work can focus on incorporating error compensation steps into the remanufacturing framework, such as error compensation for scanning errors and RE modeling errors. A detailed future comparative study of only applying direct mesh printing versus incorporating each additional error compensation step in the remanufacturing process can provide insights into the significance of various error compensations. Such systematic analysis can lead to more refined guidelines for remanufacturing workflow with RE and AM.

### 4.3 Feature sensitivity in compensation for RE in AM

For RE and remanufacturing tasks where achieving net shape is crucial, surface machining or post-processing options are unavailable, and/or stress concentration avoidance and assembly fit are necessary, distortion compensation becomes vital. Our findings indicate that features prone to process-induced distortions, scanning errors, and/or CAD generation errors – such as internal features, overhangs, fillets, and vertical holes – require careful attention. Additionally, the extent of distortions and errors varies depending on the size and features of the part, as observed in the differing deviation distributions for the two case study parts. Hence, considering part size and features is crucial when conducting part compensation for RE and rapid remanufacturing. Iterative geometric feature-based compensation, as the methods presented here, enables more precise adjustments to specific features, potentially leading to more accurate modifications. However, identifying and resolving these features may require extra manual effort. Prioritizing functionally critical features (e.g., hole diameter, slot length) by applying weights to more important ones can improve accuracy for complex geometries and features.

While this study demonstrated the compensation framework using simplified geometries, the proposed CAD-based compensation approach is inherently scalable to more complex geometries. Scaling to complex geometries involves systematically identifying additional KDCs through automated feature recognition and iterative refinement based on distortion simulation feedback. Future work can focus on integrating advanced geometric feature extraction algorithms to streamline the identification and compensation of KDCs in highly complex geometries. Future efforts could also address existing limitations by automating feature identification, integrating designer input, and streamlining technological connections, further enhancing the effectiveness and practicality of the proposed approaches.

### 4.4 STL-based vs. CAD-based compensation performance

Comparing the STL and CAD-based compensation, the CAD-based method showed a slightly lower average compensation error than the STL-based method. One contributing factor could be the uniform application of the compensation factor across all parts in the STL-based method, which leads to certain KDCs diverging during iterations. For instance, in the STL-based approach, the overhang specimen exhibits a higher compensation error due to divergence in KDCs, such as hole diameter and overall height, during the final iteration. In the perforated thin plate, the width and slot length also diverge significantly in the final iteration, resulting in the highest compensation errors for the STL-based approach. However, the CAD-based method still struggles to compensate for some specific features, which can be attributed to those features' rough surfaces and the accuracy of distortion prediction tools. Both compensation approaches rely on finite element-based distortion predictions, which can present challenges, particularly in cases involving complex material compositions, phase changes, and intricate geometries (e.g., sharp and rounded corners) (Afazov *et al.*, 2021). Here, fillets, holes, overhangs, heights, and slots exhibited higher distortion during processing, leading to more significant surface irregularities and deviations between the generated CAD model and the scanned part, causing more compensation errors. Comparative analysis indicates that the perforated thin plate experiences higher errors, possibly due to its size, which can introduce additional process-induced measurement, CAD generation, and compensation-related errors into the process.

Another limitation of the STL-based approach is its output format, which provides the final compensated model as an STL file. Since STL files are not easily modified, further adjustments to the model are restricted. Additionally, STL output may contain tessellation errors, presenting problems for printing (Peter *et al.*, 2020). Furthermore, although the input geometry was a flat parametric CAD model, the compensated file was a rough STL representation, unable to preserve the flat features of a parametric CAD model accurately. This discrepancy is evident in the comparison between the scan of the remanufactured part from the STL approach and the original model, and in the images of the remanufactured parts (Figure 2, Figure 4b, and Figure 4e). Significant deviations in the flat surfaces, such as the side and front surfaces for the overhang test specimen and the top, side, and front surfaces for the perforated thin plate, are observed.

The CAD-based approach requires running one simulation for every iteration. In contrast, the STL-based approach necessitates running two simulations per iteration (one for finding the compensated model and another for predicting the distorted output part using the compensated model as input). Based on the convergence plot (Figure 3), additional iterations in the STL-based approach may decrease the APE for most KDCs, while increasing it for a few. Applying the same compensation factor for all features may affect convergence for critical features. Similarly, the CAD-based approach does not ensure convergence for all features (Figure 3), but fewer features diverge, and the amount of divergence is lower than that of the STL-based approach.

### 4.5 Design guidelines and a hybrid compensation approach

Our results can guide part designers attempting to RE a metallic component and remanufacture using AM. Applying the same compensation factor to the entire part could increase errors for certain features. Therefore, identifying the most critical features of the part and prioritizing their convergence in the process will improve outcomes. Features highly distorted by the manufacturing process and exhibiting non-linear behavior should receive greater attention. For future work, a hybrid approach combining STL-based and CAD-based methods may yield better results, as the CAD-based method maintains the flatness of certain features, while the STL-based method provides better compensation for part lengths compared to the CAD-based approach. Moreover, future work could extend the convergence strategy by incorporating criteria based on the rate of change of APE between iterations, rather than relying solely on its absolute value, to provide a more robust stopping condition. Based on our results, for AM part compensation, selection of significant KDCs should simultaneously focus on geometric features that directly influence part functionality while also prioritizing features that are highly sensitive to LPBF process parameters, such as temperature, laser power, or build orientation, or are prone to distortion (e.g., holes, overhangs, thin features) (Gao *et al.*, 2024; Idriss *et al.*, 2018; Maleki *et al.*, 2025). Our results (as presented in Figure 4) indicate that overhang surfaces, curved features such as fillets and holes, and thin features such as part thickness, are particularly susceptible to process-induced distortion and scanning error. Designers should prioritize these surfaces as KDCs, especially if they serve an important function or are used as datums for other features.

### 4.6 End-to-end error modeling in RE and remanufacturing

The RE and remanufacturing process involves multiple interdependent stages—3D scanning, model reconstruction, distortion simulation, and geometric compensation—each introducing uncertainty. These uncertainties propagate and accumulate across the workflow, impacting the geometric accuracy of the remanufactured parts. While our framework integrates scanning and simulation-based iterative compensation to address cumulative distortion, it does not explicitly model the error propagation and interactions between individual RE stages.

Recognizing this gap, future work could focus on developing end-to-end predictive models that explicitly represent the entire RE–AM process as a sequence of state transitions. In such a representation, both geometry and associated errors systematically evolve from the initial physical scan through CAD reconstruction and iterative compensation stages, ultimately leading to the final remanufactured component. To facilitate this comprehensive modeling, advanced approaches such as state-space modeling and stream-of-variation analysis will be highly valuable.

State-space modeling would allow researchers and practitioners to track geometric deviations and their sources at each stage, providing detailed insight into how upstream errors influence downstream stages. Similarly, stream-of-variation analysis can quantify how variability introduced at earlier process stages propagates, impacting downstream geometric outcomes. These methodologies will enable precise, stage-specific error attribution, predictive analysis, and adaptive geometric compensation strategies informed by identified upstream error sources. Such advancements will enhance the fidelity, reliability, and robustness of future RE and AM workflows.

# 5 CONCLUSION

This study addressed critical challenges in applying RE techniques to AM, particularly the significant geometric distortions induced by AM processes. Although RE has proven successful in conventional manufacturing contexts, its adaptation for AM presents unique complications due to intricate part geometries and pronounced process-induced distortions. To overcome these issues, we developed an integrated, unified RE–AM framework that explicitly incorporates 3D scanning, CAD model reconstruction, and iterative AM process-simulation-driven geometric compensation. Specifically, our framework proposed and comparatively evaluated two key dimensional characteristics (KDC)-based compensation approaches—STL-based and CAD-based—leveraging iterative AM simulations to compensate for process-induced distortions.

Two case studies demonstrated the utility and comparative effectiveness of these methods. The CAD-based approach consistently outperformed the STL-based approach in maintaining critical geometric features and surface fidelity, achieving an average APE as low as 0.087%, compared to 1.82% for the STL-based method. We found that the STL-based compensation particularly struggled to maintain surface flatness and accurately compensate for certain intricate features such as holes and fillets, resulting in significant dimensional deviations in these regions. Additionally, our results underscored the impact of measurement errors, notably in small and intricate geometries, on the accuracy of the overall remanufacturing workflow.

A key contribution of our work is establishing the foundational structure for a unified and distortion-aware RE–AM pipeline. This integrated approach contrasts with traditional workflows that treat scanning, CAD reconstruction, and geometric compensation as separate, disconnected stages. Our workflow explicitly acknowledges interdependencies across stages, enhancing coherence and dimensional accuracy. While this study compensates for accumulated geometric deviations at the final stage, it does not explicitly quantify or attribute observed geometric deviations to individual stages within the RE–AM workflow, such as scanning, CAD reconstruction, and fabrication. A future step will involve stage-aware modeling, such as state-space modeling and stream-of-variation analysis, to systematically analyze and minimize error propagation throughout the workflow. These approaches can systematically capture and represent how dimensional errors accumulate across RE stages, enabling precise error attribution and adaptive compensation strategies directly informed by upstream error sources. Similarly, this work employs fixed scaling factors for iterative dimension adjustments to demonstrate the feasibility of the proposed framework. However, future work can extend this baseline by integrating gradient-informed optimization techniques and regression-based surrogate models to enhance convergence efficiency and robustness, while reducing computational cost.

Furthermore, while our study primarily demonstrated compensation strategies using simplified geometries and single, straightforward CAD modeling parameters (e.g., overall length or width), practical manufacturing contexts frequently involve highly complex geometries and KDCs resulting from combinations of multiple interacting CAD parameters. To address these practical complexities, future research should systematically integrate advanced geometric feature extraction algorithms for automating KDC identification and iterative refinement based on distortion simulation feedback. Future efforts should incorporate parametric sensitivity analysis and multi-dimensional optimization algorithms to effectively handle KDCs that inherently depend on multiple interacting CAD parameters. In doing so, techniques such as Jacobian-based sensitivity mapping and principal component analysis (PCA) of deviation data can be particularly valuable for revealing underlying parameter interdependencies. Leveraging these approaches can guide the formulation of joint compensation strategies that effectively address coupled sources of error, thereby enhancing both geometric accuracy and functional performance in reverse-engineered parts.

In summary, our integrated framework demonstrates substantial promise in leveraging simulation and scanning data to enhance the geometric accuracy and reliability of AM remanufacturing processes. It emphasizes the importance of a holistic view of the RE–AM process rather than treating each stage independently. By systematically addressing current limitations through future stage-aware error modeling, advanced feature extraction, and multi-parameter compensation strategies, our work provides a foundation and clear pathways toward even more precise, robust, and reliable additive manufacturing outcomes. This study, therefore, represents an essential first step toward a unified and distortion-aware RE–AM framework, with significant potential for industrial remanufacturing scenarios where original design data are unavailable.

## APPENDIX

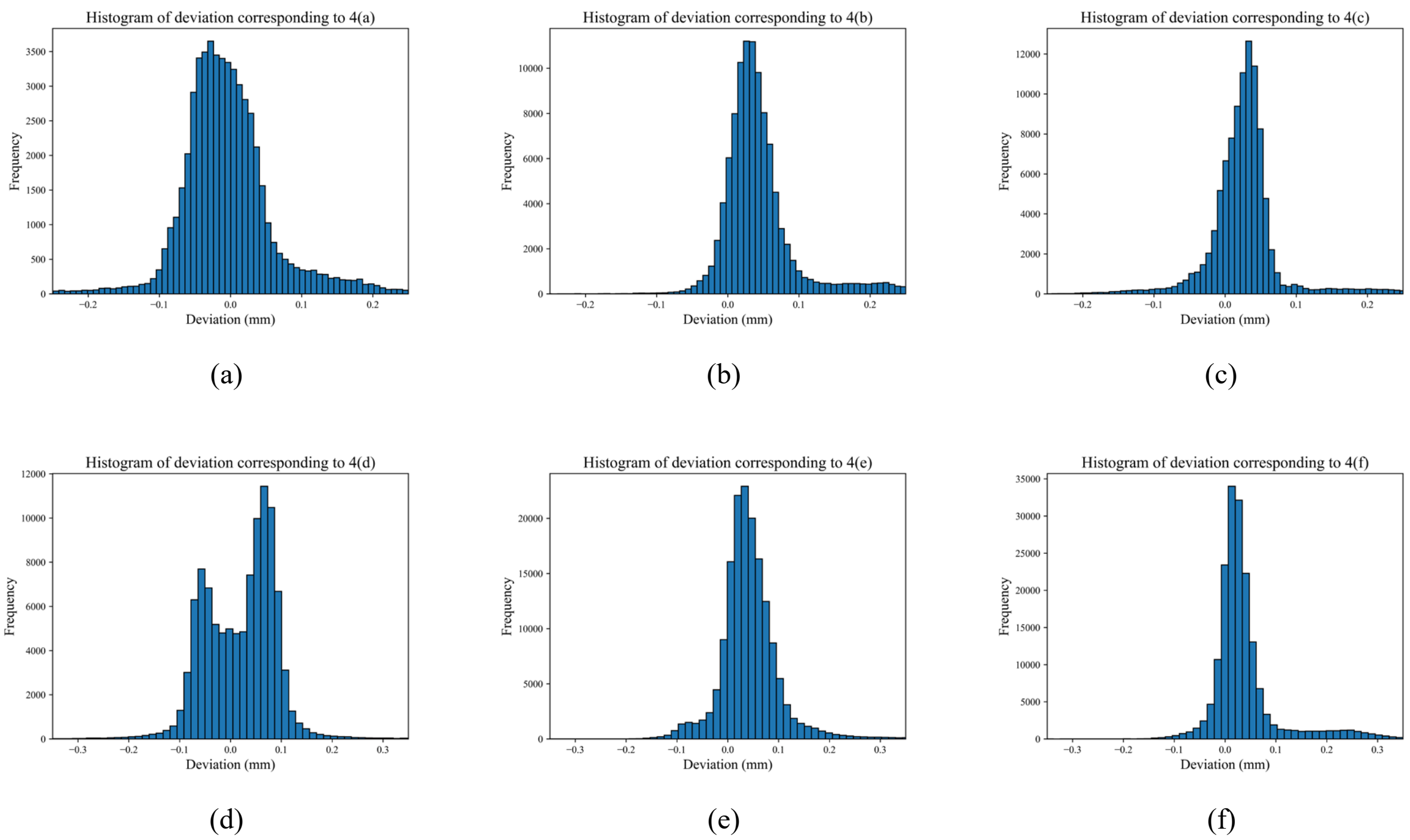

**Figure 6** Histograms of deviations for the best fit alignment of the original model and scan of (a, d) manufactured part, (b, e) remanufactured part produced using STL method, (c, f) remanufactured part produced using CAD method for the overhang test specimen and the perforated plate.